\documentstyle[prd,epsfig,aps,draft]{revtex}

\begin{document}
\twocolumn[\hsize\textwidth\columnwidth\hsize\csname
@twocolumnfalse\endcsname
\sloppy  
\draft
\title{Brane Worlds and the Cosmic Coincidence Problem}
\author{Massimo Pietroni}
\address{{\it
INFN -- Sezione di Padova,\\
Via F. Marzolo 8, I-35131 Padova, Italy}}
\maketitle
\begin{abstract}{
Brane world models with `large' extra dimensions with radii in the 
$r_l \sim 10- 100\;\mu{\rm m}$ range and smaller ones at  
$r_s \leq (1 \,{\rm TeV})^{-1}$ have the potential to solve the cosmic 
coincidence problem, {\it i.e.} the apparently fortuitous equality between
dark matter and dark energy components today.
The main ingredient is the assumption of a stabilization mechanism fixing
the total volume of the compact submanifold, but allowing for
shape deformations.
The latter are associated with phenomenologically safe 
ultra-light scalar fields.
Bulk fields Casimir energy  naturally plays the role of dark 
energy, which decreases in time because
of expanding $r_l$. Stable Kaluza Klein states may play the role
of dark matter with increasing, $O(1/r_s)$, mass. 
The cosmological equations exhibit attractor solutions in which 
the global equation of state is 
negative, the ratio between dark energy and dark matter is constant and 
the observed value of the ratio is obtained for two large extra 
dimensions.

Experimental searches of large extra dimensions should take into account that,
 due to the strong coupling between dark matter and radii dynamics, the size 
of the large extra dimensions inside the galactic halo may be smaller than
 the average value.

}
\end{abstract}
DFPD/02/TH/06,

\vskip 2pc]
\def\beqra{\begin{eqnarray}}

\def\eeqra{\end{eqnarray}}

\def\beq{\begin{equation}}

\def\eeq{\end{equation}}

\def\ds{\displaystyle}

\def\ts{\textstyle}

\def\ss{\scriptstyle}

\def\sss{\scriptscriptstyle}

\def\Vb{\bar{V}}

\def\phb{\bar{\phi}}

     \def\rhb{\bar{\rho}}

    \def\L{\Lambda}

    \def\T{\Theta}

\def\re#1{(\ref{#1})}

        \def\D{\Delta}

       \def\G{\Gamma}

\def\p{\partial}

\def\half{\mbox{\small$\frac{1}{2}$}}  

 \def\de{\delta}



\def\lta{\mathrel{\vcenter{\hbox{$<$}\nointerlineskip\hbox{$\sim$}}}}
\def\gta{\mathrel{\vcenter{\hbox{$>$}\nointerlineskip\hbox{$\sim$}}}}
\renewcommand{\Re}{\mathop{\mathrm{Re}}}
\renewcommand{\Im}{\mathop{\mathrm{Im}}}
\newcommand{\tr}{\mathop{\mathrm{tr}}}
\newcommand{\Tr}{\mathop{\mathrm{Tr}}}
%


%

\def\mm{\rm mm}
\def\mic{\mu {\rm m}}
\def\i{i}

 \def\f{f}

 \def\d{d}

 \def\e{e}

\section{Introduction}

The energy content of the Universe is nowadays, more than ever, the
subject of intense research. The picture emerging is the
following \cite{parameters}; 
the total energy density of the  Universe equals the
critical value  $\rho_c \simeq (2 \cdot 10^{-3}\, {\rm eV})^4 \simeq
(100\, \mic)^{-4}$  ({\it i.e.} $\Omega \simeq 1$),
 non-relativistic dark matter (DM) accounts for at 
most forty percent of it ( $\Omega_{DM}\simeq 0.2 - 0.4$), 
and the rest is in the form of a smooth - {\it i.e.}
non-clusterized - component, named Dark Energy (DE) (plus a small amount
 of baryons, $\Omega_b \simeq 0.05$).
Luminosity-redshift measurements of supernovae Ia \cite{supernovae}, 
as well as the
resolution of the so called age  problem \cite{age},
agree with this picture and point towards a DE with negative  equation
of state $W=p/\rho$, where $p$ and $\rho$ are pressure and energy
density, respectively.  A viable DE candidate is then Einstein's
cosmological constant ($W=-1$), but more general fluids  with $W<0$
have been considered,
emerging, for instance,  from the dynamics of scalar fields with
appropriate potentials.  The old cosmological constant problem
\cite{Wein} now becomes double-faced.  On one hand we still have the
old puzzle of the cosmological constant -- or the  scalar field
potential-- being so tiny compared to any common particle physics
scale. On the other hand -- and this is the new face of the problem-- 
the presence of DE today
in roughly the same amount as DM energy density is 
embarassing. 
Indeed, the ratio between the two energy  densities
scales as
\[
\frac{\rho_{DM}}{\rho_{DE}}\sim a^{3 W}\;,
\] 
where $a$ is the scale factor of the Universe. So, the approximate
equality between the two  components today looks quite an amazing
coincidence in the cosmic history if DE has $W\neq 0$  as required
by observations.

In principle, brane world scenarios offer a suggestive contact with the DE
problem. Indeed, extra spatial  dimensions compactified to a size as
large as $100 \mic$ have been shown to be a  viable
possibility, provided that no Standard Model (SM) field propagates through
them \cite{arkani}. 
It might then  be tempting to attribute the observed value of
the DE energy density to the Casimir energy  associated to some field
propagating in these large extra dimensions of size $r$, {\it i.e.}
$\rho_{DE} \simeq B/r^4$, with $B$ a constant typically $O(10^{-5}-10^{-4})$ 
(see sect.~IV).

If the radius $r$ is stabilized, then the
energy density behaves  exactly as a cosmological constant, and we
have made no substantial progress with respect to any of the two problems
mentioned above. Indeed, we have no clue about why the Casimir energy
would be the only non-vanishing contribution to the cosmological
constant and, moreover, the equality between the latter and the matter
energy density would appear as fortuitous as before.

In order to gain some insight at least on the cosmic coincidence
problem it seems then necessary to make the radius
$r$ dynamical, so that it may evolve on a cosmological time-scale. 
But this turns out to be quite dangerous. 
The `radion' field, whose expectation value fixes $r$,
couples to the trace of the SM energy momentum tensor  with gravitational
strength. Moreover, in order to
evolve with $\dot{r}/r \sim H$ ($H$ being the Hubble parameter), it must be 
extremely light, $m \simeq H_0 \simeq 10^{-33}\;{\rm eV}$. It then behaves
as a massless Jordan-Brans-Dicke scalar with $O(1)$ couplings to matter,
whereas present bounds are $O(10^{-3})$ \cite{will}.
Thus, the geometric explanation of a dynamical DE looks quite unlikely 
(see however \cite{albrecht}).

The purpose of this paper is to show that the above conclusion is
indeed premature, and that standard brane world scenarios with large extra 
dimensions may quite naturally solve the cosmic coincidence problem.
Unfortunately, we will add nothing
new with respect to the first problem, that of the vanishing of all 
present contributions to the
cosmological constant larger than $O(100\, \mic)^{-4}$.

The key point in our discussion is the realization that  besides 
large extra-dimensions in the $10-100 \,\mic$ range there could also be
smaller ones. Considering for simplicity 
only two compact subspaces, each caracterized by a single radius, we note that
the trace of the 
four-dimensional
energy-momentum tensor couples with the total {\it volume} of the compact 
space, that
is with the combination $r_s^{n_s} r_l^{n_l}$, where $r_{s,l}$ and $n_{s,l}$ 
are the radii and dimensionalities of the two subvolumes, 
respectively. 
If we assume a
stabilization mechanism that fixes the {\it total volume} ${\cal V}$
 of the compactified
manifold but {\it not its shape}, the potentially dangerous combination of 
radion fields associated
with volume fluctuations is made harmless, whereas the orthogonal one, 
associated with
shape deformation, is decoupled from normal matter and may then 
be ultra-light.
As a consequence, $r_l$ can grow on a cosmological
time-scale (such that the associated Casimir energy $B/r_l^4$
decreases) and at the same time $r_s$ shrinks so that ${\cal V}$ keeps a 
fixed value.

The coincidence problem is solved if we
associate the dark matter with some stable state living in the extra 
dimensions, because the resulting
cosmological  equations exhibit an {\it attractor solution} in which the
Casimir energy and $\rho_{DM}$ are redshifted at the
same rate with their ratios fixed at a $O(1)$ value. 

If the DM candidate has a mass
independent on $r_s$ then the attractor corresponds to the equation of
state of non-relativistic  matter, {\it i.e.} $W=0$, which is
disfavoured by cosmological observations.

On the other hand, if the mass of the DM particle is inversely proportional 
to some power of $r_s$, as for a Kaluza-Klein (KK)
state ($m_{DM} = m_{KK} \sim r_s^{-1}$), then
a more interesting behavior is obtained. Indeed, on  the attractor now we have
\[
\rho_{DM} \sim 1/(r_s a^3) \sim \rho_{DE} ;
\;,
\]
which, due to the shrinking of $r_s$, decreases slower than $a^{-3}$
during the cosmic expansion.  Remembering that a fluid with equation
of state $W$ scales as $\rho\sim a^{-3(W+1)}$ we see that the energy densities
scale now with a  {\it negative} effective
equation of state. Moreover, 
the values of $W$, $\Omega_{DM}$, and $\Omega_{DE}$ depend  only
on the two parameters, $n_s$ and $n_l$. 

The above results rely on naive
dimensional reduction, without assuming drastic modifications to the
radion  kinetic terms as was done for instance in \cite{albrecht}.  

For sake of clarity, we summarize our assumptions here. They are;

1) the vanishing of any contributions to the effective 4-dimensional 
cosmological constant
larger than $O(100 \,\mic)^{-4}$ today;

2) a stabilization mechanism for the total volume (rather than for each single
 radius separately)
of the compactified space effective at low  energies.

The paper is organized as follows. In sect.~II we introduce our setting and 
derive the effective action after dimensional reduction. In sect.~III 
we discuss a mechanism to stabilize the volume of the compact manifold, 
allowing at the same time its shape to vary. In sect.~IV we will study the 
dynamics of the 
modulus associated to shape deformations of the compact manifold, showing 
that an (attractor) solution exists on which dark energy and dark matter scale
at the same ratio and the universe has a negative effective equation of state.
Finally, in sect.~V we discuss our results.

\section{The setting}
The starting point is the action
\beq
S=S_{\rm bulk} + S_{4+n_s} + S_4\;.
\eeq
The bulk action is given by
\beq
S_{\rm bulk} =\int d^{4+{\cal{N}}} X 
\sqrt{-\cal G}
\left[\frac{R({\cal G_{A B}})}{16\pi G_{4+{\cal N}}} + \Lambda_B+ \ldots\right]\;,
\label{sbulk}
\eeq
where ${\cal N}=n_s+n_l$ and the $4+{\cal N}$- dimensional metric is given by
\beqra
&&\ds
d s^2 = {\cal G}_{A B} d X^A d X^B \nonumber \\
&&= g_{\mu\nu} dx^\mu dx^\nu + \sum_{i=s,l} b_i(x)^2 
\gamma^{(i)}_{\alpha_i \beta_i} dy^{\alpha_i} dy^{\beta_i}\;,
\eeqra 
where $\mu,\nu =0,\ldots,3$, $\alpha_i, \beta_i = 1,\ldots, n_i$, and 
$b_i(x)=r_i(x)/r_i^0$, with $r_i^0$'s the average values of the radii today. 
$\Lambda_B$ is the bulk cosmological constant, whereas the dots in eq. (\ref{sbulk}) represent extra fields living in the bulk which will contribute to the stabilization of the compact volume, as we will discuss in the next section.

Apart from the
overall $b_i$'s factors, 
the metrics $\gamma^{(i)}$'s are assumed to be non-dynamical.

$S_{4+n_s}$ is the action for the fields living in all the
 $4+n_s$ dimensions. In the following, we will only assume
that there is at least one stable KK state, which we will associate to DM.

The SM fields are confined to our four-dimensional brane, with action given by 
\beq
S_4=\int d^4x\sqrt{-g}\left[{\cal L}_{SM}(g_{\mu\nu}, \psi) +  
\Lambda_4
\right]\;,
\eeq
where $\psi$ represent SM fields and we have singled out the contributions to the cosmological constant from the brane, $\Lambda_4$.
The dimensional reduction of $S_{\rm bulk}$ is a straightforward 
procedure (see for instance \cite{carroll}). We define new
`radion' variables, $\phi_a(x)$, as
$\ln b_i = A_{ia}\phi_a$
with $i= s,l$, $a=1,2$, and
\beq
A = \frac{1}{\sqrt{\cal N}} 
\left[
\begin{array}{cc}
\sqrt{\frac{2}{2+{\cal N}}} & \sqrt{\frac{n_l}{n_s}}\\
\sqrt{\frac{2}{2+{\cal N}}} & -\sqrt{\frac{n_s}{n_l}}
\end{array}
\right]\,.
\label{Amatrix}
\eeq
By also rescaling 
the 4-dimensional metric as 
\beq
g_{\mu\nu} = \e^{-C_{\cal N} \phi_1} \tilde{g}_{\mu\nu}\,,
\label{einstein}
\eeq
where $C_{\cal N}=\sqrt{2 {\cal N}/(2+{\cal N})}$,
we obtain the `Einstein frame' action,
\beqra
&& \ds S_{\rm bulk} = \int d^4x \sqrt{-\tilde{g}} \left[ \frac{1}{16 \pi G_4}
\tilde{R}(\tilde{g}_{\mu\nu}) \right. \;\;\;\;\;\;\;\;\;\;\nonumber \\ 
&& \ds \left. - \frac{M_p^2}{2} \sum_{a=1}^2 \tilde{g}^{\mu\nu}\partial_\mu
\phi_a \partial_\nu \phi_a +\e^{-C_{\cal N} \phi_1} {\cal V} \Lambda_B + \ldots \right]\;, \;\;\;\;\;\;\;\;\;\;
\label{sbulk_e}
\eeqra
where we assumed that both $R(\gamma^{(i)})=0$, as in toroidal 
compactifications, and ${\cal V} = G_{4+{\cal N}}/G_4$ is the volume of the compact manifold today.

In the same frame, $S_4$ takes the form
\beqra
S_4= && \ds
\int d^4x \sqrt{-\tilde{g}} \e^{-2 C_{\cal N} \phi_1} 
\left[{\cal L}_{SM}(\e^{- C_{\cal N} \phi_1} \tilde{g}_{\mu\nu}, \psi) 
\right.\nonumber\\
&&\ds \left.+ \Lambda_4 \right]\;,
\label{s4_e}
\eeqra
from  which we see that only $\phi_1$, which we might call `the volumon', 
couples to matter. Indeed, from eq.~(\ref{Amatrix}) one realizes that $\phi_1$ is the scalar field controlling volume deformations, since
\beq
e^{C_{\cal N} \phi_1} = b_s^{n_s} b_l^{n_l}\,.
\eeq


The orthogonal combination, $\phi_2$, controls shape deformations of the compact manifold and does not couple to SM fields directly. 
On the other hand, it couples to fields living in the bulk, or on the $4+n_s$-dimensional
brane. In the
4-dimensional
language, $\phi_2$ couples -- besides the graviton-- 
to {\it non-zero} KK states, since their mass terms are proportional 
to $1/r_i \sim \exp(-A_{i2}\phi_2)$  ($i=s,l$).
 
\section{Volume stabilization}
The field $\phi_1$ couples to the SM as a Jordan-Brans-Dicke scalar, {\it i.e.} in a purely metric way. As is well known, this type of couplings enforces the (weak) equivalence principle, but is constrained from gravity tests in the solar system and in the laboratory \cite{will}.
A massless $\phi_1$  would modify gravity at an unacceptable level unless $C_{\cal N} < O(10^{-3})$, which is clearly ruled out in the present model, as $C_{\cal N}$ ranges from 1 ($n_s = n_l = 1$) to $\sqrt{2}$ ($n_s + n_l \to \infty $). It is then necessary to give a mass to $\phi_1$, that is, to stabilize the volume of the compact manifold. In order to fulfill the experimental bounds, the $\phi_1$ range should be smaller than about a hundred microns, that is, $m_{\phi_1} \gta 10^{-3} {\rm eV}$. At the same time, we would like to keep $\phi_2$ massless such that it may play the role of dynamical dark energy today, that is, $m_{\phi_2} \lta H_0 \sim 10^{-33} {\rm eV}$. Thus, we are looking for a stabilization mechanism which fixes only the total volume of the compact manifold, allowing the different dimensions to vary accordingly.

A mechanism with this property has been discussed for instance in refs.~\cite{SADM}, and is based on the introduction of a $U(1)$ gauge monopole configuration in the extra dimensions, besides the bulk cosmological constant $\Lambda_B$ and the `brane tension', $\Lambda_4$, that we have already introduced. 

For definitness, we will first consider a two dimensional torus of radii $b_l$ and $b_s$, and then we will discuss the extension of the mechanism to higher dimensional manifolds. The key point is that $U(1)$ gauge fields can have non-zero magnetic flux through the two-dimensional surface, and the flux is quantized and topological invariant, 
\beq \Phi \equiv \int d^2y \, \epsilon^{m n} F_{m n} = 2 \pi k\;,
\label{flux}
\eeq
where $m,n = 4,5$, $\epsilon^{45}=-\epsilon^{54} =1$, and $k$ is an integer. Adding the gauge kinetic term,
\beq \int d^{4+2}X \sqrt{-{\cal G}} \frac{M_\star^2}{4g^2} F_{A B}F^{A B}
\label{gkin}
\eeq
($A,B = 0,\ldots,5$), to $S_{\rm bulk}$ and solving the Euler-Lagrange equations for $F_{A B}$, a solution can be found of the form
\[
F_{\mu \nu}=F_{\mu m} =0\;,\;\;\;\;\;F_{m n}= \frac{\epsilon_{m n} B}{b_l b_s}\;,\]
with $\partial_m B=0$. $M_\star$ in (\ref{gkin}) is a mass scale typically $O(G_6^{-1/4})$.
Inserting this solution in eq.~(\ref{flux}) we find that flux quantization forces the electromagnetic field to be inversely proportional to the 2-dimensional volume,
\[B=\frac{k \pi}{{\cal V} b_l b_s}\;,\] so that eq.~(\ref{gkin}) becomes
\beq
\frac{k^2 \pi^2 M_\star^2}{2 g^2 {\cal V}} \int d^4 x\sqrt{-\tilde{g}} e^{-3 C_2 \phi_1}\;.
\label{sgauge_e}
\eeq
Like the terms proportional to $\Lambda_{B}$ in eq.~(\ref{sbulk_e}), and to $\Lambda_4$ in eq.~(\ref{s4_e}), eq.~(\ref{sgauge_e}) is only sensitive to the volume, that is to $\phi_1$, and not to the shape modulus, $\phi_2$. These contributions give the effective potential for $\phi_1$ in the Einstein frame, which is the appropriate quantity to minimize in order to study stabilization\footnote{In the second of refs.~\cite{SADM} they minimized the effective potential in the frame with metric $g_{\mu\nu}$, where the field $\phi_1$ appears also in front of the Ricci scalar, which should be taken in into account in the minimization. As a consequence, their results on the mass of the scalar field differ significantly from ours.},  {\it i.e}
\beq
V(\phi_1) = D e^{- C_2 \phi_1} + A e^{- 3 C_2 \phi_1} + \Lambda_4 e^{- 2 C_2 \phi_1}\;,
\eeq
with \[ 
A=\frac{k^2 \pi^2 M_\star^2}{2 g^2 {\cal V}}\;,\;\;\; D =  {\cal V} \Lambda_B\,.\]
Notice that $A>0$, which ensures that the potential is bounded from below. Requiring that the contribution of the potential to the cosmological constant vanishes in the minimum, we get the fine tuning condition
\[ \Lambda_4^2 = 4 A D\,,\] which forces $D>0$, and then $\Lambda_B>0$. The potential has a minimum only if $\Lambda_4<0$, that is, for $\Lambda_4=-2\sqrt{A D}$, in 
\[ \phi_1^{\rm min} = \frac{1}{C_2} \log \sqrt{\frac{A}{D}}\;.\]

On this minimum, the scalar field $\phi_1$ has a mass
\beqra
  m^2_{\phi_1} &=& V''(\phi_1^{\rm min})/M_p^2\\
&=& 
\frac{2^{3/2} C_{2}^2}{k \pi} (r_s^0 M_\star)^2 (r_l^0 M_\star)^2 \left(\frac{\Lambda_B}{M_\star^6}\right)^{3/2} \frac{M_\star^4}{M_p^2}\;.
\eeqra

The above mechanism can be generalized to compact spaces of higher dimensions. In order to stabilize the volume of a ${\cal N}$-dimensional manifold, one needs a 
${\cal N}-1$-form $U(1)$ gauge field, with ${\cal N}$-form field strength $F^{(\cal N)}$. Then, the flux (\ref{flux}) generalizes to the topological invariant 
\[ M_\star^{{\cal N} -2} \int_{\cal M_ N} F^{(\cal N)}= 2 \pi k \,,\]
and the kinetic energy is again inversely proportional to the compact volume \cite{SADM}. 
Considering again toroidal compactifications, the ${\cal N}$-dimensional case gives a scalar mass
\beq 
m^2_{\phi_1} =\frac{2^{3/2} C_{\cal{N}}^2}{k \pi} (r_s^0 M_\star)^{2n_s} (r_l^0 M_\star)^{2 n_l} \left(\frac{\Lambda_B}{M_\star^{4+{\cal N}}}\right)^{3/2} \frac{M_\star^4}{M_p^2}\;.
\eeq
The bulk cosmological constant is bounded from above by the requirement that it does not induce a bulk curvature radius smaller than $r_l^0$ \cite{SADM}, which gives
an upper bound on $m_{\phi_1}$,
\beq 
m^2_{\phi_1} \lta \frac{2^{3/2} C_{\cal{N}}^2}{\pi} (r_s^0 M_\star)^{3 n_s/n_l} \left(\frac{M_p}{M_\star}\right)^{4-6/n_l}\frac{M_\star^4}{M_p^2}\;,
\eeq where we have taken $k=1$. Assuming $r_s M_\star \gta 1$ and $M_\star \gta 10 {\rm TeV}$ we see that the requirement $m_{\phi_1}\gta 10^{-3} {\rm eV}$ is in no conflict with the bound above for any $n_l \geq 2$.

Since $m_{\phi_1} > H$ ($H$ being the Hubble parameter) 
for temperatures $T \lta M_\star$, in the following we will consider the field $\phi_1$ fixed during the late time evolution of the Universe, and will study the dynamics of the light $\phi_2$ field. When $\phi_1$ is constant and massive, the two frames related by eq.~(\ref{einstein}) are equivalent.

\section{Vamps from extra-dimensions}

Integrating the non-zero KK modes out leaves `Casimir'
contributions to the free 
energy proportional to $1/r_{s,l}^4$ . For instance, a scalar field 
compactified on a circle of radius $r$  with periodic (antiperiodic) boundary conditions gives a contibution $-5.06\cdot 10^{-5}/r^4$ ($4.74\cdot 10^{-5}/r^4$)  \cite{Candelas}, whereas a fermion gives $2.02\cdot 10^{-4}/r^4$ ($-1.90\cdot 10^{-4}/r^4$). For higher dimensional torii, the numerical coefficients change, but the order of magnitude is the same (on a two-torus with $r_1=r_2=r$, we find $-4.87\cdot10^{-5} /r^4$  ( $1.95\cdot 10^{-4}/r^4$) for a scalar (fermion) with periodic boundary conditions.

Consistently with our assumption 1) in the Introduction, we will neglect the 
$O(r_s^{-4})$ term, since it must be cancelled by
the (unknown) mechanism solving the cosmological constant problem. A possible mechanism could be supersymmetry in the bulk or, even without supersymmetry, a particle content in the $4+n_s$ brane, cancelling exactly the $1/r_s^4$ contribution, in the same spirit of ref.~\cite{Dienes}.

Thus, we are left with the Casimir contribution from the `large' dimensions,  given by
\beq V(\phi_2)= \frac{B}{r_l^4}= \frac{B}{(r_l^0)^4} 
\exp\left(4 \sqrt{\frac{n_s}{n_l {\cal N}}} \phi_2\right)\,,
\label{potential}
\eeq
where we have fixed $\phi_1=0$ and
$B$ is a $O(10^{-4} - 10^{-5})$ coefficient depending on the particle content and the dimensionality of the
$4+{\cal N}$-dimensional bulk, which we require to be positive.

As we have anticipated, the other ingredient of our model is dark matter. 
We associate it to a stable KK state, whose mass scales as $O(1/r_s)$. It will be an example of varying mass dark matter particles, or Vamps, which 
were discussed in a different context in refs.~\cite{vamp}.

The important point is that the cosmological abundance of this 
non-relativistic relic will scale in this case as
\beq
\rho_{DM} \sim r_s^{-1} a^{-3} \sim 
\exp\left(- \sqrt{\frac{n_l}{n_s {\cal N}}}
 \phi_2\right) a^{-3}\,.
\label{dm}
\eeq
Since the runaway potential, eq.~(\ref{potential}), pushes the field 
$\phi_2$ to
$-\infty$, the mass of the  dark-matter particle increases during the 
cosmological expansion, and its energy density redshifts less than for common 
dark matter.

The fact that $\phi_2$ is the canonically normalized version of a radion field
is the reason for the exponential dependences in (\ref{potential}) and 
(\ref{dm}). This is of crucial importance for what follows, because 
exponentials, once inserted in the cosmological equations, allow {\it scaling}
solutions, in which $\rho_{DE}$ and $\rho_{DM}$ redshift at the same rate.

Defining
\beq
\beta= 4 \sqrt{\frac{n_s}{n_l {\cal N}}} \,,\;\;\;\;\;
\lambda= \sqrt{\frac{n_l}{n_s {\cal N}}}\;,
\eeq
the cosmological equations are,
\beqra
&&\ds \ddot{\phi_2} + 3 H \dot{\phi_2} = - \beta \,V + \lambda \,\rho_{DM}
\nonumber\\
&&\ds 
H^2 = \frac{1}{3 M_p^2} \left(\rho_{DM} +\frac{M_p^2}{2}\dot{\phi_2}^2 +V
\right)
\nonumber\\
&&\frac{\ddot{a}}{a}= - \frac{1}{6 M_p^2}
\left( \rho_{DM} +2 M_p^2\dot{\phi_2}^2 -2 V \right)\;.
\label{friedmann}
 \eeqra
They admit a solution of the form 
\[ \phi_2 = \frac{-3}{\lambda+\beta} \log a \;, \]  
($a_0=1$)
which is such that $\rho_{\phi_2} \sim \rho_{DM} \sim a^{-3(W+1)}$, with the 
equation of state
\beq
W= \frac{-\lambda}{\lambda+\beta} =  
- \left(1 + 4 \frac{n_s}{n_l} \right)^{-1}\,,\;
\label{eos}
\eeq
and fixed ratio,
\beqra
\Omega_{DE} &&=\ds \frac{\rho_{\phi_2}}{ \rho_{DM} +\rho_{\phi_2}} = 
\frac{3 + \lambda(\beta +\lambda)}{(\beta+\lambda)^2}\,,\nonumber\\
&&=\frac{n_l ( 3 n_s {\cal N} + 4 n_s  + n_l )}{(4 n_s + n_l )^2}\;,
\eeqra
independent on the scale factor $a$. 
The solution above is an attractor over solution space if 
\beq
n_s > \frac{n_l (3 n_l - 4 )}{16 - 3 n_l}\;.
\label{attractor}
\eeq
Differently from the case of a 
cosmological constant, or even from quintessence models with inverse power 
law potentials, once the attractor is reached $\Omega_{DE}$ and $\Omega_{DM}$
 become 
independent on the cosmological era, thus solving the cosmic coincidence 
problem. Moreover, the equation 
of state (\ref{eos}) is negative, as required by observations.

The values of $W$ and $\Omega_{DE}$ on the attractor are 
functions of $n_s$ and $n_l$ only.
In Tab.~1 we list the possible values
of $\Omega_{DE}$, $W$ and of 
$H_0 t_0$, $t_0$ being the present age of the Universe. 
We limited the dimensionality of
the compact space according to the theoretical prejudice coming from string 
theory, {\it i.e.} ${\cal N} \le 6$.

\begin{quote}
\begin{tabular}{||c|c|c|c|c||}
\hline
$n_s$ & $n_l$ & $\Omega_{DE}$ & $W$ & $H_0 t_0$\\
\hline
\hline
1 - 5 &   1 &      $\le 0.44$  &  --  & -- \\
\hline
1     &   2 &  0.83            & $-1/3$ & 1 \\
\hline
2     &   2 &  0.68            &  -0.20& 0.83\\
\hline
3     &   2 &  0.60            &  -0.14& 0.78\\
\hline
4     &   2 &  0.56            & -0.11 & 0.75\\
\hline
1-3   &   3 &  $\ge 0.92$      &  --     &   --  \\
\hline
1-2   &   4 & no attractor& -- & -- \\
\hline
1    &5& no attractor & -- & -- \\
\hline
\hline
\end{tabular}  
\end{quote}
\begin{itemize}
\item[Tab.1] The values of $\Omega_{DE}$, $W$, and $H_0 t_0$ for 
different values of $n_s$ and $n_l$ such that ${\cal N}=n_s+n_l\le 6$.
\end{itemize}

\section{discussion}

The first noticeable fact about Tab.~1 is that the observed range for
the dark energy density, $0.6 \lta \Omega_{DE} \lta 0.8$ 
uniquely selects the number of `large' extra dimensions to be $n_l=2$, the
same value that is required by the totally unrelated issue of solving the
hierarchy problem with `millimeter' size extra dimensions \cite{arkani}.
 
Indeed, at the level of Tab.~1, we have not yet inserted any information
about absolute scales, such as $100 \,\mic$, TeV, $H_0$ and so on. We find it
a remarkable and inspiring `coincidence' that the observed balance between
$\Omega_{DM}$ and $\Omega_{DE}$ points to the same value,
$n_l=2$, obtained from scale dependent considerations, such
 as reproducing $H_0$ or
solving the hierarchy problem.
For  $n_l\neq 2$ we either find that the attractor corresponds to energy
densities outside the observed range or that the couplings $\lambda$ and 
$\beta$ lie outside the limit of eq.~(\ref{attractor}) and correspond to a 
different --unphysical --  attractor.

The values for $H_0 t_0$ listed in the table are obtained from
\[
H_0 (t_0 -t_{\rm att})  = \frac{2}{3(W+1)} 
\left( 1- a_{\rm att}^{\frac{3}{2}(W+1)}\right)\;,
\]
where $t_{\rm att}$ and $a_{\rm att}$ are the time and scale factor when the 
attractor is reached, assuming $a_{\rm att} < 10^{-2}$.
Taking  the $95\%$ 
confidence level lower limit on $t_0$ from globular cluster age estimates, 
$t_0 > 11 {\rm Gyr}$ \cite{age} we obtain the lower bound 
\[
H_0 t_0 > (0.71 - 0.79)
\]
 where we have varied $H_0$ in the range $(63 - 70)\,{\rm km \,s^{-1} 
Mpc^{-1}}$. The above bound 
is inconsistent with a flat, matter-dominated universe (for which 
$W=0$ and $H_0 t_0 = 2/3$) while it is satisfied by all the 
$n_l=2$ models in Tab.~1.

Supernovae Ia data, taken at face value, point towards a more negative equation
of state than those listed in Tab.~1, typically in the $W\lta -0.6$ range 
\cite{supernovae}. 
However, the analyses have been done assuming two fluids with
different equations of state, {\it i.e.}
matter ($W=0$) and `quintessence' ($W=W_x$), whereas in the present case the
two fluids scale with the same, negative, equation of state. 
 Since quintessence begins to dominate
the energy density quite recently, the negative equation of state takes over 
later than in our model, where it has been negative since much previous 
epochs. As a consequence, the supernovae bounds on the equation
of  state of the universe should be somehow relaxed in our model compared to 
quintessence.

Structure formation in a generic model of the type of eq.~(\ref{friedmann}) 
was studied in \cite{amendola} (where also the baryonic component was 
considered). There, it was shown that the non-zero -- and constant-- 
$\Omega_{DM}$ allows 
the growth of perturbations even if the expansion is accelerated. This behavior
is to be contrasted with the usual quintessence or cosmological constant case,
where the perturbations freeze out soon after the takeover because in that 
case $\Omega_{DM}$ drops quickly to zero.

The model we have presented in this paper is just the simplest version of a 
large family. In order to discuss its phenomenological implications thoroughly,
one should specify it in detail. Different dependences of the dark matter mass
on the small radius could also be envisaged, possibly leading to different 
predictions for the equation of state. Also, the couplings of the
KK sector to the SM are crucial. If the dark matter candidate is not a SM 
singlet, then its mass variation would induce variations in the gauge coupling
constants. Imposing the existing strong constraints on the latter
would imply that the $\phi_2$ field was frozen up to a not too distant 
past and that the 
attractor became effective quite recently.

It is, anyway, not trivial at all that the 
simplest choices, that is, tree-level dimensional reduction
 and the assumption of a 
$O(10^{-4}/r_l^4)$ Casimir energy, lead to a model which solves the cosmic 
coincidence problem and is --if not fully compatible with-- at least very
close to the observations. 
A non-trivial point to consider is the persistence of the behavior of the 
tree-level solution once 
radiative corrections are included. The issue can be treated consistently only
once a detailed model is given, however we refer the reader to the discussion
in ref.~\cite{albrecht}, where supersymmetry in the bulk is argued to cancel 
all dangerous corrections larger than $O(1/(r_l^0)^2 M_p)$ to the radion mass.

A model-independent prediction of the framework outlined
in this paper is the presence of extra 
dimensions in the $10-100\, \mic$ range. Whereas for the solution of the hierarchy
problem discussed in \cite{arkani} this value was {\it allowed}, the link 
with the cosmological expansion discussed in this paper makes it a true
{\it prediction}. The expected value 
for $r_l$ may be quite close to the present bound of
$200 \,\mic$, 
obtained from measurements of the Newton's law at small distances 
\cite{eotwash}. However, one cautionary remark is due on this point. 
The predicted value is determined by global 
observables, {\it i.e.} $H_0$ and $\ddot{a_0}/a_0$. Since the radius and DM 
are strongly coupled in this model, it is plausible that the local value of 
$r_l$, inside the galactic halo, would be somehow different. 
It is difficult to estimate the size of this effects, which would require a 
dedicated numerical study of the halo formation in this model. However, the 
sign would plausibly go in the direction of shrinking $r_l$ inside the halo. 
This is because the system would use the additional degree of freedom 
represented by the varying mass, to decrease the gravitational potential in 
overdense regions, thus expanding $r_s$ --and then shrinking $r_l$ -- with
respect to the average universe.

The effect of varying mass dark particles on cosmic perturbations and on 
the structure of halos will be 
the subject of a forthcoming publication \cite{MPS}.
\\

\vspace{0.5 cm}
It is a pleasure to thank Tony Riotto for useful discussions. This work was 
partially supported by European Contracts HPRN-CT-2000-00148 and
HPRN-CT-2000-00149.


\begin{thebibliography}{99}
\bibitem{parameters} For a review see, for instance:
M.~Fukugita and C.~J.~Hogan,
Eur.\ Phys.\ J.\ C {\bf 15}, 136 (2000).

\bibitem{supernovae} 
A.~G.~Riess {\it et al.}  [Supernova Search Team Collaboration],
Astron.\ J.\  {\bf 116}, 1009 (1998)
S.~Perlmutter {\it et al.}  [Supernova Cosmology Project Collaboration],
Astrophys.\ J.\  {\bf 517}, 565 (1999)

\bibitem{age} 
L.~M.~Krauss and B.~Chaboyer,
astro-ph/0111597.

\bibitem{Wein} 
S.~Weinberg,
Rev.\ Mod.\ Phys.\  {\bf 61}, 1 (1989).

\bibitem{arkani} 
N.~Arkani-Hamed, S.~Dimopoulos and G.~R.~Dvali,
Phys.\ Lett.\ B {\bf 429}, 263 (1998),
Phys.\ Rev.\ D {\bf 59}, 086004 (1999);
I.~Antoniadis, N.~Arkani-Hamed, S.~Dimopoulos and G.~R.~Dvali,
Phys.\ Lett.\ B {\bf 436}, 257 (1998).

\bibitem{will}
C.~M.~Will,
Living Rev.\ Rel.\  {\bf 4}, 4 (2001)

\bibitem{albrecht}
A.~Albrecht, C.~P.~Burgess, F.~Ravndal and C.~Skordis,
astro-ph/0107573.

\bibitem{carroll}
S.~M.~Carroll, J.~Geddes, M.~B.~Hoffman and R.~M.~Wald,
hep-th/0110149.


\bibitem{SADM} 
R.~Sundrum,
Phys.\ Rev.\ D {\bf 59}, 085010 (1999);
N.~Arkani-Hamed, S.~Dimopoulos and J.~March-Russell,
Phys.\ Rev.\ D {\bf 63}, 064020 (2001).

\bibitem{vamp}
J.~A.~Casas, J.~Garcia-Bellido and M.~Quiros,
Class.\ Quant.\ Grav.\  {\bf 9}, 1371 (1992);
G.~W.~Anderson and S.~M.~Carroll,
astro-ph/9711288.

\bibitem{Candelas}
P.~Candelas and S.~Weinberg,
Nucl.\ Phys.\ B {\bf 237}, 397 (1984).

\bibitem{Dienes}
K.~R.~Dienes,
Nucl.\ Phys.\ B {\bf 611}, 146 (2001)

\bibitem{amendola} 
L.~Amendola and D.~Tocchini-Valentini, astro-ph/0111535.

\bibitem{eotwash}
C.~D.~Hoyle {\it et al.} Phys. Rev. Lett. {\bf 86}, 1418 (2001).

\bibitem{MPS} S.~Matarrese, M.~Pietroni and C.~Schimd, in preparation.

\end{thebibliography}
\end{document}